\documentstyle[galley, rotate]{mn}
\onecolumn
\input{epsf}
\def\n{{\noindent}}

\def\inc{{\int_0^{\chi_s}}}

\def\kmin{{\int_0^{\chi_s} \omega(\chi) d \chi }}
\def\corr{\Big [ \int  {d^2 {\bf l} \over (2
\pi)^2 )} {\rm  P}( { \bf l \over r(\chi) })  W^2(l\theta_0) \exp [ i l
\theta_{12}] \Big ]}
\def\var{ \Big [  \int  {d^2 {\bf l} \over (2
\pi)^2 )} {\rm P} ( { \bf l \over r(\chi) })  W^2(l\theta_0) \Big ] } 
\def\one{\langle \kappa_s^2 \rangle}
\def\two{\langle \kappa_s(\gamma_1) \kappa_s(\gamma_2) \rangle_c}

\title[Bias and Weak Lensing]
{Probing The Gravity Induced Bias with Weak Lensing:\\
Test of Analytical results Against Simulations}

\author[Dipak Munshi]{Dipak Munshi \\
Max-Planck-Institut fur Astrophysik,
Karl-Schwarzschild-Str.1, D-85740, Garching, Germany\\
}
\pagerange{000,000}

\begin{document}

\maketitle

\begin{abstract}
Future weak lensing surveys will directly
probe the density fluctuation in the universe. Recent studies have
shown how the statistics of the weak lensing convergence field 
is related
to the statistics of collapsed objects. Extending earlier analytical
results on the probability distribution function of the convergence 
field we show that the bias associated with the convergence field can 
directly be related to the bias associated with the statistics of 
underlying over-dense objects. This will provide us a direct method 
to study the gravity induced bias in galaxy clustering. Based on our
analytical results which use the hierarchical {\em ansatz} for 
non-linear clustering, we study how such a bias depends on the 
smoothing angle and the source red-shift. We compare our analytical 
results against ray tracing experiments through N-body simulations 
of four different realistic cosmological scenarios
and found a very good match. Our study shows that the bias
in the convergence map strongly depends on the background geometry 
and hence can help us in distinguishing different cosmological models
in addition to improving our understanding of the gravity 
induced bias in galaxy clustering.       
\end{abstract}


\section{Introduction}

Weak gravitational lensing (Bartelmann \& Schneider 1999) is the 
distortion of images of the
background galaxies through the tidal gravitational field of
large scale background mass distribution. Study of such distortions
provides us an unique way to probe statistical properties of 
intervening mass distribution.    

Although there is no generally applicable definition of weak lensing
the common aspect of all studies in weak gravitational lensing is that
measurements of its effects are statistical in nature.
Pioneering  works in this direction were done by Blandford et
al. (1991), Miralda-Escude (1991) and Kaiser (1992) based on the 
earlier work by Gunn (1967).  Current progress in weak lensing
can broadly be divided into two distinct categories. Where as Villumsen
(1996), Stebbins (1996), Bernardeau et al. (1997) and Kaiser (1987)
have focussed on linear and quasi-linear regime by assuming a 
large smoothing angle, several authors have developed a numerical 
technique to simulate weak lensing catalogs. Numerical
simulations of weak lensing typically employ N-body simulations,
 through which ray tracing experiments are conducted
(Schneider \& Weiss 1988; Jarosszn'ski et al. 1990; Lee \& Paczyn'ski
1990; Jarosszn'ski 1991; Babul \& Lee 1991;  Bartelmann \& Schneider
1991, Blandford et al. 1991). Building on the earlier work of Wambsganns
et al. (1995, 1997, 1998) most detailed numerical study of lensing 
was done by  Wambsganns, Cen \& Ostriker (1998). Other recent studies
using ray tracing experiments have been conducted by Premadi, Martel
\& Matzner (1998), van Waerbeke, Bernardeau \& Mellier (1998),
Bartelmann et al (1998) and Couchman, Barber \& Thomas (1998). While 
peturbative analysis can provide valuable information at large
smoothing angle such analysis can not be used to study lensing 
at small angular scales as the whole perturbative series starts to
diverge. 

A complete analysis of the weak lensing statistics at small angular
scales will require a similar analysis
for the underlying dark matter distribution which we do not have at 
present. However there are several 
non-linear {\em ansatze} which predict a tree hierarchy for 
matter correlation functions and are thought to be successful to 
some degree in modeling
results from numerical simulations. Most of these {\em ansatze}
assumes a tree hierarchy for higher order correlation functions and 
they disagree with each other by the way they assign weights to
trees of same order but of different topologies (Balian \& Schaeffer 1989,
Bernardeau \& Schaeffer 1992; Szapudi \& Szalay 1993). Evolution of
two-point correlation functions in all such approximations 
are left arbitrary. However recent studies by several authors 
(Hamilton et al 1991; Nityananda \& Padmanabhan 1994; Padmanabhan et
al. 1996; Jain, Mo \& White 1995; Peacock \& Dodds 1996) have provided us a
very accurate fitting formula for the
evolution of the two-point correlation function which can be used in
combination with these hierarchical {\em ansatze} to predict
clustering properties of dark matter distribution in the universe.

Most recent studies in weak lensing have mainly focussed on 
lower order cumulants (van Waerbeke, Bernardeau \& Mellier 1998,
Hui 1999, Munshi \& Jain 1999a), 
cumulant correlators (Munshi \& Jain 1999b) and errors 
associated with their measurement from observational data using 
different filter functions (Reblinsky et
al. 1999). However it is well known that higher order 
moments and more sensitive to the tail of the distribution function which they
represent and are sensitive to measurement errors due to 
finite size of catalogs (Szapudi \& Colombi 1996). On the other hand numerical simulations 
involving ray tracing techniques have already shown that while 
probability distribution function (PDF) associated with density field is
not sensitive to cosmological parameters its weak lensing counterpart 
can lift such a degeneracy and can help us in estimation of cosmological parameters
from observational data (Jain, Seljak \& White 1999). 
Munshi \& Jain (1999a,b) extended such studies to show that the
hierarchical {\em ansatz} can actually be used to make concrete analytical
predictions for lower order statistical properties of the convergence field.
Valageas (1999a) has
used hierarchical {\em ansatz} for computing error involved in 
estimation of $\Omega_0$ and $\Lambda$ from
SNeIa observations. A similar fitting function was recently proposed by
Wang (1999). Our formalism is similar to that of Valageas (1999a)  
although the results obtained by us include the effect of 
smoothing and analysis presented here correspond to the joint
probability distribution function of two smoothed patches and hence
the bias associated with them. Our
formalism presented here is an extension of the study of the PDF using
similar techniques (Munshi \& Jain  1999b, Valageas 1999b), it can very easily be extended
to multi-point PDF and hence to compute the $S_N$ parameters 
associated with ``hot spots'' of the convergence maps. Analytical results
and  detailed comparison against numerical simulations will be presented
elsewhere (Munshi \& Coles 1999a,b).

The paper is organized as follows. In section $2$ we describe briefly
ray tracing simulations. Section $3$ presents most of the analytical 
results necessary for computation of the bias of the smoothed 
convergence field $\kappa_s$. Results of data analysis are 
presented in section $4$. Section $5$ is left for discussion of 
our result in general cosmological framework.

\section{Generation of Convergence Maps from N-body Simulations}

 Convergence maps are generated by solving the geodesic equations for
the propagation of light rays through N-body simulations of dark matter
clustering. The simulations used for our study are adaptive 
$P^3M$ simulations with
$256^3$ particles and were carried out using codes kindly made
available by the VIRGO consortium. These simulations can resolve
structures larger than $30h^{-1}kpc$ at $z = 0$ accurately. These
simulations were carried out using 128 or 256 processors on CRAY T3D
machines at Edinburgh Parallel Computer Center and at the Garching
Computer Center of the Max-Planck Society. These simulations were
intended primarily for studies of the formation and clustering of
galaxies (Kauffmann et al 1999a, 1999b; Diaferio et al 1999) but were
 made available by these authors and by the Virgo
Consortium for this and earlier studies of gravitational lensing 
(Jain, Seljak \& White 1999, Reblinsky et al. 1999, Munshi \& Jain
1999a,b).

Ray tracing simulations were carried out by Jain et al. (1999) using
a multiple lens-plane calculation which implements the discrete
version of recursion relations for mapping the photon position and the
Jacobian matrix (Schneider \& Weiss 1988; Schneider, Ehler \& Falco
1992). In a typical experiment $4\times 10^6$ rays are used to trace 
the underlying mass distribution. The dark matter distribution 
between the source and the observer is projected onto $20$ - $30$ planes.
The particle positions on each plane are interpolated onto a $2048^2$ 
grid. On each plane the shear matrix is computed on this grid 
from the projected density by using Fourier space relations between the
two. The photons are propagated starting from a rectangular grid on the
first lens plane. The regular grid of photon position gets distorted 
along the line of sight. To ensure that all photons reach the
observer, the ray tracing experiments are generally done backward in time
from the observer to the source plane at red-shift $z = z_s$. 
The resolution of the convergence maps is limited by 
both the resolution scale associated with numerical simulations and also 
due to the finite resolution of the grid to propagate photons. 
The outcome of these simulations are shear and 
convergence maps on a two dimensional grid. Depending on the 
background cosmology the two dimensional box 
represents a few degree scale patch on the sky. For more details on the
generation of $\kappa$-maps, see Jain et al (1999).

\section{Analytical Predictions}
In this section we will provide necessary theoretical background for
the bias function in the context of hierarchical clustering.
 
We will be using the following line element for the
background geometry of the universe:
\begin{equation}
d\tau^2 = -c^2 dt^2 + a^2(t)( d\chi^2 + r^2(\chi)d^2\Omega)
\end{equation}
Where we have denoted angular diameter distance by $r(\chi)$ and
scale factor of the universe by $a(t)$. $r(\chi)= K^{-1/2}\sin (K^{-1/2}
\chi)$ for positive curvature, $r(\chi) = (-K)^{-1/2}\sinh
((-K)^{-1/2}\chi)$ for negative curvature and $\chi$ for the flat
universe. For a present value of  of $H_0$
and $\Omega_0$ we have $K= (\Omega_0 -1)H_0^2$. The various parameters 
charecterising different cosmological models are listed in Table - 1.

\begin{table}
\begin{center}
\caption{Cosmological Parameters charecterising different models}
\label{tabsig2D}
\begin{tabular}{@{}lcccc}
                  &SCDM&TCDM&LCDM&OCDM\\
$\Gamma$&0.5&0.21&0.21&0.21\\
$\Omega_0$&1.0&1.0&0.3&0.3 \\
$\Lambda_0$&0.0&0.0&0.7&0.0 \\
$\sigma_8$&0.6&0.6&0.9&0.85\\
$H_0$&50&50&70&70\\
\end{tabular}
\end{center}
\end{table}

\subsection{The Formalism}

The statistics of weak lensing
convergence $\kappa$ is very much similar to that of the projected
density field. In what follows we will be considering a small patch
of the sky where we can use the plane parallel approximation or small
angle approximation to replace spherical harmonics by Fourier modes.
The 3D density contrast $\delta$ along the line of sight when projected into 2D
sky with a weight function $\omega(\chi)$ will provide us the
projected density contrast or the weak-lensing convergence at a
direction $\gamma$.

\begin{equation}
\kappa({\bf \gamma}_1) = \inc {d\chi}_1
\omega(\chi_1)\delta(r(\chi){\bf \gamma}_1)
\end{equation}

\n
In all our discussion  we will be placing  the sources at a fixed
red-shift (an approximation not too difficult to modify for more
realistic description), the weight function can be expressed as $\omega(\chi)
= 3/4a c^{-2}H_0^2 \Omega_m r(\chi) r(\chi_s - \chi)/ r(
\chi_s)$. Where $\chi_s$ is the comoving radial distance to the source
at a redshift $z_s$. Fourier decomposition of $\delta$ can be written as:

\begin{equation}
\kappa(\gamma_1) = \inc {d\chi}_1 \omega(\chi_1) \int {d^3{\bf k} \over {(2
\pi)}^3} \exp ( i \chi_1 k_{\parallel} + i r \theta k_{\perp} ) \delta_k
\end{equation}

\n
Where we have used $k_{\parallel}$ and $k_{\perp}$ to denote components
of wave vector ${\bf k}$ parallel and perpendicular to line of sight
direction $\gamma$. In small angle approximation however one assumes
that $k_{\perp}$ much larger compared to $k_{\parallel}$. We will denote the 
angle between the line of sight direction ${\bf \gamma}$ and the wave
vector ${\bf k}$ by $\theta$. Using the definitions we have introduced
above we can compute the smoothed
projected two-point correlation function (Peebles 1980, Kaiser 1992, Kaiser 1998):

\begin{equation}
\langle \kappa(\gamma_1) \kappa(\gamma_2) \rangle_c = \inc d {\chi_1}
{\omega^2(\chi_1) \over r^2(\chi_1)} \int {d^2 {\bf l} \over (2
\pi)^2}~\exp ( \theta l )~ {\rm P} { \big ( {l\over r(\chi)} \big )}
 W_2^2(l\theta_0).
\end{equation}

\n
Where we have introduced a new notation ${\bf l} = r(\chi){\bf
k}_{\perp}$ which denotes a scaled wave vector projected on the
surface of the sky. The average of two-point correlation function
$\langle \kappa_s^2 \rangle$ 
smoothed over an angle $\theta_0$ with a top-hat smoothing window
$W_2(l\theta_0)$ is useful to quantify the fluctuations in
$\kappa_s$ which is often used to reconstruct the matter power
spectrum $P({\bf k})$ (Jain, Selzak \& White 1998).

\begin{equation}
\langle \kappa_s^2 \rangle_c = \inc d {\chi_1}
{\omega^2(\chi_1) \over r^2(\chi_1)} \int {d^2 {\bf l} \over (2
\pi)^2}~ {\rm P} { \big ( {l\over r(\chi)} \big )} W_2^2(l\theta_0)
\end{equation}

Similar analysis for the higher order cumulant correlators (Szapudi \&
Szalay 1997, Munshi \& Coles 1999) of the
smoothed convergence
field relating $\langle \kappa_s^m (\gamma_1) \kappa_s^n (\gamma_2) \rangle_c$ with
multi-spectra of underlying dark matter distribution $B_p$ (Munshi \& 
Coles 1999a):

\begin{equation}
 \langle \kappa_s^2(\gamma_1) \kappa_s(\gamma_2)
\rangle_c =
 \int_0^{\chi_s} { \omega^3 (\chi) \over r^4(\chi) } d \chi \int
 \frac{d^2{\bf l}_1}{(2\pi)^2} \int  \frac{d^2{\bf l}_2}{(2\pi)^2}  W_2(l_1
 \theta_0) W_2(l_2 \theta_0) W_2( l_3\theta_0) \exp(il_2
 \theta_{12}) B_3 \left( {{\bf l}_1 \over r (\chi)},
 {{\bf l}_2 \over r (\chi)}, {{\bf l}_3 \over r (\chi)} \right);
\end{equation}

\begin{eqnarray}
\langle \kappa_s^3(\gamma_1) \kappa_s(\gamma_2) \rangle_c =
\int_0^{\chi_s} { \omega^3 (\chi) \over r^4(\chi) } d \chi \int
 \frac{d^2{\bf l}_1}{(2\pi)^2} \int  \frac{d^2{\bf l}_2}{(2\pi)^2}  \int  \frac{d^2{\bf
 l}_3}{(2\pi)^2} W_2(l_1 \theta_0) W_2(l_2 \theta_0) W_2( l_3\theta_0)W_2( l_4\theta_0) \exp(il_3 \theta_{12})
  \nonumber\\
 B_4 \left( {{\bf l}_1 \over r (\chi)},
 {{\bf l}_2 \over r (\chi)}, {{\bf l}_3 \over r (\chi)}, {{\bf l}_4
\over r (\chi)} \right);
\end{eqnarray}

\begin{eqnarray}
\langle \kappa_s^2(\gamma_1) \kappa_s^2(\gamma_2) \rangle_c =
\int_0^{\chi_s} { \omega^3 (\chi) \over r^4(\chi) } d \chi \int
 \frac{d^2{\bf l}_1}{(2\pi)^2}  \int \frac{d^2{\bf l}_2}{(2\pi)^2}  \int  \frac{d^2{\bf
 l}_3}{(2\pi)^2}
W(l_1 \theta_0) W_2(l_2 \theta_0) W_2( l_3\theta_0)W_2( l_4\theta_0)
\exp(i(l_1+l_2)\theta_{12})
\nonumber  \\ 
 B_4 \left( {{\bf l}_1 \over r (\chi)},
 {{\bf l}_2 \over r (\chi)}, {{\bf l}_3 \over r (\chi)}, {{\bf l}_4
\over r (\chi)} \right).
\end{eqnarray}

\n
In general we can express the cumulant correlators of arbitrary order 
$\langle \kappa_s^m(\gamma_1) \kappa_s^n(\gamma_2) \rangle_c$ in terms
of multi-spectra $B_{m+n}$ as:

\eject

\begin{eqnarray}
\langle \kappa_s^m(\gamma_1) \kappa_s^n(\gamma_2) \rangle_c =
\int_0^{\chi_s} { \omega^{n + m} (\chi) \over r^{2(n+m-1)}(\chi) }
d \chi \int \frac{d^2 {\bf  l}_1}{(2\pi)^2} \dots  \int
\frac{d^2{\bf l}_{n+m-1}}{(2\pi)^2} W_2(l_1 \theta_0)\dots W_2(
l_{n+m}\theta_0) \exp[i(l_1 + \dots + l_m)\theta_{12}] \nonumber \\
 B_{m+n} \left( {{\bf l}_1 \over r (\chi)},
 \dots, {{\bf l}_{m+n} \over r (\chi)} \right).
\end{eqnarray}

\n
We will use and extend these results in this paper to show that it is possible
to compute the whole bias function
$b(>\kappa_s)$, i.e. the bias associated with those spots in 
convergence map which $\kappa_s$ is above certain threshold (which acts as a generating function for these 
cumulant correlators) from the statistics of underlying over-dense dark objects.Details of analytical results presented here can be found in 
Munshi \& Coles (1999b).

\subsection{Hierarchical {\em Ansatze}}

In deriving the above expressions we have not used any specific form
for the
matter correlation hierarchy, however the length scales involved 
in small angles are in the highly non-linear regime. Assuming a tree model 
for the  matter correlation
hierarchy in the highly non-linear regime one can write the most
general case as (Groth \& Peebles 1977; Fry \& Peebles 1978;  Davis \&
Peebles 1983; Bernardeau \& Schaeffer 1992; Szapudi \& Szalay 1993):

\begin{equation}
\xi_N( {\bf r_1}, \dots {\bf r_N} ) = \sum_{\alpha, \rm N-trees}
Q_{N,\alpha} \sum_{\rm labellings} \prod_{\rm edges (i,j)}^{(N-1)}
\xi({\bf r_i}, {\bf r_j})
\end{equation}

It is interesting to note that an exactly similar hierarchy
develops in the quasi-linear regime in the limit of vanishing variance
(Bernardeau 1992), however the hierarchal amplitudes $Q_{N, \alpha}$
become shape dependent in such a case. In the highly nonlinear 
regime there are some indications that these functions become
independent of shape parameters as has been proved by studies of
lowest order parameter $Q_3 = Q$ using high resolution numerical
simulations (Sccociamarro et al. 1998). In the Fourier space such an
{\em ansatz} will mean that the whole hierarchy of multi-spectra $B_p$ 
can be written in terms of sum of products of power-spectra, e.g. in
low orders we can write:

\begin{eqnarray}
&&B_2({\bf k}_1, {\bf k}_2, {\bf k}_3)_{\sum k_i = 0} = Q ( P({\bf
k_1})P({\bf k_2}) + P({\bf k_2})P({\bf k_3})
+ P({\bf k_3})P({\bf k_1}) ), \\ 
&&B_3({\bf k}_1, {\bf k}_2, {\bf k}_3, {\bf k}_4)_{\sum k_i = 0} = R_a
P({\bf k_1})P({\bf k_1 +
k_2}) P({\bf k_1 + k_2 + k_3})  + {\rm cyc. perm.} + R_b P({\bf
k_1})P({\bf k_2})p({\bf k_3}) + 
{\rm cyc. perm.} 
\end{eqnarray}

\n
Different hierarchal models differ the way they predict the
   amplitudes of different tree topologies. Bernardeau \&
Schaeffer (1992) considered the case where amplitudes in general are
factorizable, at each order one has a new ``star'' amplitude 
 and higher order ``snake'' and ``hybrid'' amplitudes are
constructed from lower order ``star'' amplitudes (see Munshi,
Melott \& Coles 1999a,b,c for a detailed description). In models proposed by
Szapudi \& Szalay (1993) it is assumed that all hierarchal amplitudes of a
given order are actually degenerate.

We do not use any of these specific models for clustering and only
assume a hierarchal nature for the higher order correlation functions. 
Galaxy surveys have been used to study these {\em ansatze}. Our main 
motivation here is to show that weak-lensing surveys can also provide 
valuable information in this direction, in addition to constraining
 matter power-spectra and the background geometry of the universe. 
The most
general form for the lower order cumulant correlators in the large
separation limit can be expressed as:

\begin{eqnarray}
\langle \kappa_s^2(\gamma_1) \kappa_s(\gamma_2) \rangle_c & = &
2Q_3 {\cal C}_3 [\kappa_{\theta_0} \kappa_{\theta_{12}}] =
C_{21}^{\eta}{\cal C}_3 [\kappa_{\theta_0} \kappa_{\theta_{12}}] \equiv C_{21}^{\kappa} \langle
\kappa_s^2 \rangle_c \langle \kappa_s(\gamma_1) \kappa_s(\gamma_2) \rangle_c, \\
\langle \kappa_s^3(\gamma_1) \kappa_s( \gamma_2) \rangle_c & = &
(3R_a + 6 R_b){\cal C}_4 [\kappa_{\theta_0}^2 \kappa_{\theta_{12}}] =
 C_{31}^{\eta}{\cal C}_4 [\kappa_{\theta_0}^2 \kappa_{\theta_{12}}]
 \equiv  C_{31}^{\kappa} \langle
\kappa_s^2 \rangle_c^2 \langle \kappa_s(\gamma_1) \kappa_s(\gamma_2) \rangle_c
,\\  \langle \kappa_s^2(\gamma_1) \kappa_s^2(\gamma_2) \rangle_c & =
& 4 R_b{\cal C}_4 [\kappa_{\theta_0}^2 \kappa_{\theta_{12}}] 
= C_{22}^{\eta}{\cal C}_4 [\kappa_{\theta_0} \kappa_{\theta_{12}}] 
\equiv  C_{22}^{\kappa} \langle
\kappa_s^2 \rangle_c^2 \langle \kappa_s(\gamma_1) \kappa_s(\gamma_2) \rangle_c ,\\
\langle \kappa_s^4(\gamma_1) \kappa_s(\gamma_2)\rangle_c & = &
(24S_a + 36S_b + 4 S_c){\cal C}_5 [\kappa_{\theta_0}^3
\kappa_{\theta_{12}}] = 
C_{41}^{\eta} {\cal C}_5 [\kappa_{\theta_0}^3 \kappa_{\theta_{12}}] 
\equiv  C_{41}^{\kappa} \langle
\kappa_s^2 \rangle_c^3 \langle \kappa_s(\gamma_1) \kappa_s(\gamma_2)
\rangle_c
,\\ \langle \kappa_s^3(\gamma_1) \kappa_s^2(\gamma_2) \rangle_c & = &
 (12S_a + 6 S_b){\cal C}_5[\kappa_{\theta_0}^3 \kappa_{\theta_{12}}] =
C_{32}^{\eta}{\cal C}_5[\kappa_{\theta_0}^3 \kappa_{\theta_{12}}]
\equiv  C_{32}^{\kappa} \langle
\kappa_s^2 \rangle_c^3 \langle \kappa_s(\gamma_1) \kappa_s(\gamma_2)
\rangle_c.
\end{eqnarray}

\n
Where $C_{mn}^{\kappa}$ denotes the cumulant correlators of the
convergence field and $C_{mn}^{\eta}$ denotes the cumulant correlators
for the underlying mass distribution. Extending above results to arbitrary order we can write: 

\begin{equation}
 \langle \kappa_s^p(\gamma_1) \kappa_s^q(\gamma_2)\rangle_c  =
 C_{pq}^{\eta}{\cal C}_{p+q}[\kappa_{\theta_0}^{(p+q-2)} \kappa_{\theta_{12}}] 
= C_{pq}^{\kappa} \langle
\kappa_s^2 \rangle_c^{(p+q-2)} \langle \kappa_s(\gamma_1) \kappa_s(\gamma_2)
\rangle_c.
\end{equation}

\n
where $C_{pq}^{\eta}$ denotes the cumulant correlators for the
underlying mass distribution,

\begin{equation}
{\cal C}_t[\kappa^m_{\theta_0} \kappa_{\theta_{12}} ] =
\int_0^{\chi_s} { \omega^t(\chi)\over
r^{2(t-1)}(\chi)}\kappa^m_{\theta_0}\kappa_{\theta_{12}} d\chi,
\end{equation}

\n
similarly the following notations were used to simplify the above expressions:

\begin{eqnarray}
\kappa_{\theta_0} & \equiv & \int  \frac{d^2\bf l}{(2\pi)^2} P
\left( {l \over r(\chi)} \right) W_2^2(l \theta_0), \\
\kappa_{\theta_{12}} & \equiv & \int
 \frac{d^2\bf l}{(2\pi)^2} P \left( {l \over r(\chi)} \right)
W_2^2(l \theta_0) \exp ( l \theta_{12}).
\end{eqnarray}.

\noindent

The hierarchical expression for the lowest order cumulant 
i.e. $S_3$ was derived by Hui 1998. He also
showed that his result agrees well with numerical ray tracing
experiments by Jain, Seljak and White (1998). More recent studies
have shown that higher order cumulants and even the
two-point statistics such as cumulant correlators 
can also be reliably modeled in a similar way (Munshi \& Coles 1999, Munshi \&
Jain 1999a). We extend such results in this paper to compute the complete 
bias function $b_{\kappa}(\kappa_s)$ (which is related to the low order
two-point statistics
such as the cumulant correlators defined above) associated with high 
$\kappa_s$ spots in the convergence map,

\begin{eqnarray}
&&p_{\kappa}(\kappa_1, \kappa_2)d\kappa_1 d\kappa_2 = p_{\kappa}(\kappa_1) p_{\kappa}(\kappa_2)( 1
+ b_{\kappa}(\kappa_1) \xi^{\kappa}_{12} b_{\kappa}(\kappa_2)) d\kappa_1 d\kappa_2,  \\ \nonumber
\end{eqnarray}

\n
and its relation to the bias associated with collapsed objects in 
underlying density field $1+\delta$.


\subsection{The bias associated with collapsed objects}

The success of analytical results in predicting the lower order 
cumulants, cumulant
correlators  and the one-point smoothed PDF(Munshi \& Coles 199a,b;
Munshi \& Jain 1999a,b; Valageas 1999a,b), motivates a general
analysis of the bias associated with the
high $\kappa_s$ spots in convergence maps and their relation
with bias associated with the high peaks in the underlying mass
distribution.  For this
purpose we found that formalism developed by Balian \&
Schaeffer (1989) and Bernardeau \& Schaeffer (1992) (later extended 
Bernardeau (1992, 1994)) to be most suitable. 
These results are based on a very  general tree hierarchy for
higher order
correlation function and the assumption that amplitudes associated 
with different tree-topologies are constant once in the highly
non-linear regime. Later these results were generalized by 
Bernardeau (1992, 1994) for the case of quasi-linear regime too, 
where perturbative dynamics can be used to make more concrete
predictions. Errors associated with top-hat smoothing and its use 
with hierarchical {\em ansatz} has been studied in detail by Szapudi 
et al.(1992) and Boschan et al. (1994) and will be assumed small even in
the case of two-point cumulant correlators throughout this paper. 
In this section we review the basic results from 
scaling models in the highly non-linear regime and quasi-linear regime
 before extending such models to the statistics of 
smoothed convergence field $\kappa(\theta_0)$ where we will show that
although weak lensing statistics probes the highly non-linear regime,
projection effects make the variance smaller than unity justifying the
use of quasi-linear results even though the generating function 
remains the one for highly non-linear regime. We will be
using the small angle approximation in our derivation. 
Our results can in principle be generalized for the case of 
projected galaxy catalogs too, a detailed analysis will be presented 
elsewhere.

\subsubsection{The Generating Function}

In scaling analysis of the probability distribution function (PDF) the
void probability distribution function (VPF) plays most fundamental 
role, which can related to the generating function of the cumulants 
or $S_N$ parameters, $\phi(y)$ (White 1979, Balian \& Schaeffer 1989) 
\begin{equation}
P_v(0) = \exp ( -\bar N \sigma(N_c) ) = \exp \Big ( - { \phi (N_c) \over
\bar \xi_2} \Big  ).
\end{equation}
\n
Where $P_v(0)$ is the probability of having no ``particles'' in a cell of 
of volume $v$, $\bar N$ is the average occupancy of these ``particles'' and 
$N_c = \bar N {\bar \xi}_2$. The VPF is meaningful only for discrete 
distribution of particles and can not be defined for smooth density 
fields such as $\delta$ or $\kappa(\theta_0)$. However the scaling
functions defined above
$\sigma(y) = -{\phi(y) \over y}$ are very much useful even for
continuous distributions where they can be used as a generating
function of one-point cumulants or $S_p$ parameters,
\begin{equation}
\phi(y) = \sum_{p=1}^{\infty} { S_p \over p! } y^p.
\end{equation}
The function $\phi(y)$ satisfies the constraint $S_1 = S_2 = 1$
necessary for proper normalization of PDF. The other generating function
which plays a very important role in such analysis is the generating 
function for vertex amplitudes $\nu_n$, associated with nodes appearing in
tree representation of higher order correlation hierarchy ($Q_3 =
\nu_2$, $R_a = \nu_2^2$ and $R_b = \nu_3$). 

\begin{equation}
{\cal G}(\tau) = 1 - \tau + { \nu_2 \over 2 ! } \tau^2 - { \nu_3 \over
3! } \tau^3 + \dots
\end{equation}

\n 
A more specific model for ${\cal G}(\tau)$ can be used, which
is useful to make more specific predictions (Bernardeau \& Schaeffer
1979):

\begin{equation} 
{\cal G}(\tau) = \Big ( 1 + {\tau \over \kappa_a} \Big )^{-\kappa_a}.
\end{equation}

\n
We will relate $\kappa_a$ with other parameters of scaling models.
While the definition of VPF do not use any specific form of
hierarchical {\em ansatz} it is to realize that writing the tree
amplitudes in terms of the weights associated with nodes is only
possible when one assumes a factorizable model for tree hierarchy
(Bernardeau \& Schaeffer 1992) and other possibilities which do not
violate the tree models are indeed possible too (Bernardeau \&
Schaeffer 1999). The generating functions for tree nodes can be 
related to the VPF by solving a pair of implicit equations 
(Balian \& Schaeffer 1989),

\begin{eqnarray}
&&\phi(y) = y {\cal G}(\tau) - { 1 \over 2} y {\tau} { d \over d
\tau} G(\tau), \\
&&\tau = -y { d \over d\tau} {\cal G}(\tau).
\end{eqnarray}

However a more detailed analysis is needed to include the effect of
correlation between two or more correlated volume element which will 
provide information about bias, cumulants and cumulant correlators 
of these collapsed object (as opposed to the cumulants and cumulant 
correlators of the whole convergence map, e.g. Munshi \& Jain (1999a)). However we will only quote results useful
for measurement of bias from ray-tracing simulations as detailed
derivations of related results including related error analysis can be
found elsewhere (Bernardeau \& Schaeffer 1992, 1999; Munshi et
al. 1999a,b,c; Coles et al. 1999).

Notice that $\tau(y)$ (also denoted by $\beta(y)$ in the literature)
plays the role of generating function for
factorized cumulant correlators $C_{p1}$ ($C_{pq} = C_{p1}C_{q1}$):

\begin{equation}
\tau(y) = \sum_{p=1}^{\infty} {C_{p1} \over p!} y^p
\end{equation}

\subsubsection {The Highly Non-linear Regime}

The PDF $p(\delta)$ and bias $b(\delta)$  can be related to their
generating functions VPF $\phi(y)$ and $\tau(y)$ respectively
by following equations (Balian \& Schaeffer 1989, Bernardeau \&
Schaeffer 1992, Bernardeau \& Schaeffer 1999),

\begin{eqnarray}
&&p(\delta) = \int_{-i\infty}^{i\infty} { dy \over 2 \pi i} \exp \Big [ {(
1 + \delta )y - \phi(y)  \over \bar \xi_2} \Big ], \label{ber1} \\ 
&&b(\delta) p(\delta) = \int_{-i\infty}^{i\infty} { dy \over 2 \pi i} \tau(y) \exp \Big [ {(
1 + \delta )y - \phi(y)  \over \bar \xi_2} \Big ] \label{ber2}. 
\end{eqnarray}

\n
It is clear that the function $\phi(y)$ completely determines 
the behavior of the PDF $p(\delta)$ for all values of $\delta$. However
different asymptotic expressions of $\phi(y)$ govern the behavior
of $p(\delta)$ for different intervals of $\delta$. For large $y$ we
can express $\phi(y)$ as:

\begin{equation}
\phi(y) = a y^{ 1 - \omega}.
\end{equation}

Where we have introduced a new parameter $\omega$ for the description of
VPF. This parameter plays a very important role in scaling analysis.
No theoretical analysis has been done so far to link $\omega$ with
initial power spectral index $n$. Numerical simulations are generally
used to fix $\omega$ for a specific initial condition. 
Such studies have
confirmed that the increase in power on smaller scales 
increases the value
of $\omega$. Typically initial power spectrum with spectral index $n=
-2$ (which should model CDM like spectra we considered in our
simulations at small length scales) produces a value
of $.3$ which we will be using in our analysis of PDF of the convergence
field $\kappa_s$ (Colombi et. al. (1992, 1994, 1995). The VPF
$\phi(y)$ and its two-point analog $\tau(y)$
both exhibit singularity for small but negative value of $y_s$,

\begin{eqnarray}
&&\phi(y) = \phi_s - a_s \Gamma(\omega_s) ( y - y_s)^{-\omega_s}, \\ \nonumber
&&\tau(y) = \tau_s - b_s ( y - y_s )^{-\omega_s - 1}.
\end{eqnarray}

\n
For the factorizable model of the hierarchical clustering the 
parameter $\omega_s$
takes the value $-3/2$ and $a_s$ and $b_s$ can be expressed in terms
of the  nature of the generating function ${\cal G}(\tau)$ and its 
derivatives near the singularity $\tau_s$ 
(Bernardeau \& Schaeffer 1992):

\begin{eqnarray}
&&a_s = {1 \over \Gamma(-1/2)}{\cal G}'(\tau_s) {\cal G}''(\tau_s) \left [
{ 2 {\cal G}'(\tau_s) {\cal G}''(\tau_s) \over {\cal G}'''(\tau_s)}
\right ]^{3/2}, \\
&&b_s = \left [
{ 2 {\cal G}'(\tau_s) {\cal G}''(\tau_s) \over {\cal G}'''(\tau_s)}
\right ]^{1/2}.
\end{eqnarray}

As mentioned before the parameter $k_a$ which we have introduced in
the definition of
${\cal G}(\tau)$ can be related to the parameters $a$ and $\omega$ appearing 
in the asymptotic expressions of $\phi(y)$ (Balian \& Schaeffer 1989, 
Bernardeau \& Schaeffer 1992),

\begin{eqnarray}
&&\omega = k_a / ( k_a + 2),\label{ka} \\
&&a = {k_a + 2 \over 2} k_a^{ k_a /  k_a + 2}.
\end{eqnarray}

Similarly the parameter $y_s$ which describe the behavior
of the function $\phi(y)$ near its singularity can be 
related to the behavior of
${\cal G(\tau)}$ near $\tau_s$ which is the solution of the equation
(Balian \& Schaeffer 1989, Bernardeau \& Schaeffer 1992),



\begin{equation}
\tau_s = {{\cal G}'(\tau_s) \over {\cal G}''(\tau_s) },
\end{equation}

\n
finally we can relate $k_a$ to $y_s$ by following expression (see eq. (\ref{ka})):

\begin{equation}
y_s = - { \tau_s \over {\cal G}'(\tau_s)},
\end{equation}

\n
or

\begin{equation}
-{ 1 \over y_s} = x_{\star} = {1 \over k_a } { (k_a + 2)^{k_a + 2} \over (k_a + 1)^{k_a+1}}.
\end{equation}

\n
The newly introduced variable $x_\star$ will be useful to define 
large $\delta$ tail of the PDF $p(\delta)$ and the bias $b(\delta)$. 
Different asymptotes  in $\phi(y)$
are linked with behavior of $p(\delta)$ for various regimes of
$\delta$. For very large values of variance i.e. $\xi_2$ 
it is possible to define a scaling function $p(\delta) = { 1 \over \xi_2^2
} h(x) $  which will encode 
the scaling behavior of PDF, where plays the role of the scaling 
variable and is defined as ${1 + \delta} \over \xi_2$. We list
different ranges of $\delta$ and specify the behavior of $p(\delta)$
and $b(\delta)$ in these regimes (Balian \& Schaeffer 1989).

\begin{equation}
{\bar \xi }^{ - \omega \over ( 1 - \omega)} << 1 + \delta << \bar \xi;
~~~~~~
p(\delta) = { a \over \bar \xi_2^2} { 1- \omega \over \Gamma(\omega)}
\Big ( { 1 + \delta \over \xi_2 } \Big )^{\omega - 2}; ~~~~~
 b(\delta) = \left ( {\omega \over 2a } \right )^{1/2} { \Gamma
(\omega) \over \Gamma [ { 1\over 2} ( 1 + \omega ) ] } \left( { 1 +
\delta \over \bar \xi_2} \right)^{(1 - \omega)/2}
\end{equation}

\begin{equation}
1+ \delta >> {\bar \xi}_2; ~~~~
p(\delta) = { a_s \over \bar \xi_2^2 } \Big ( { 1 + \delta \over \bar
\xi_2}  \Big ) \exp \Big ( - { 1 + \delta \over x_{\star} \bar \xi_2}
\Big );  ~~~~~ b(\delta) = -{ 1 \over {\cal G}'(\tau_s)} {(1 + \delta)
\over { {\bar \xi}_2}} 
\end{equation}

\n
The integral constraints satisfied by scaling function are
$ S_1 = \int_0^{\infty} x h(x) dx = 1$ and 
$ S_2 = \int_0^{\infty} x^2 h(x) dx = 1$. These take care of  
normalization of the function $p(\delta)$. Similarly the 
normalization constraint over $b(\delta)$ can be expressed as
$C_{11} = \int_0^{\infty} x b(x)h(x)dx = 1$, which translates into
$\int_{-1}^{\infty} d\delta b(\delta)p(\delta) = 0$ and
$\int_{-1}^{\infty} d\delta \delta b(\delta)p(\delta) = 1$.  
Several numerical
studies have been conducted to study the behavior of $h(x)$ and $b(x)$
for different initial conditions (e.g. Colombi et al. 1992,1994,1995; Munshi et
al. 1999, Valageas et al. 1999). For very small values of $\delta$ the behavior of
$p(\delta)$ is determined by the asymptotic behavior of $\phi(y)$ 
for large values of $y$, and it is possible to define another scaling function 
$g(z)$ which is completely determined by
$\omega$, the scaling parameter can be expressed as $z = (1+
\delta)a^{-1/(1-\omega)}{\bar \xi}_2^{\omega /(1 - \omega)}$. 
However numerically it is much easier to determine $\omega$ 
from the study of $\sigma(y)$ compared to the study of $g(z)$ 
(e.g. Bouchet \& Hernquist 1992).

\begin{equation}
1 + \delta << \bar \xi_2;~~~~
%
%
%
%
p(\delta) = a^{ -1 \over 1 - \omega} {\bar \xi}_2^{ \omega \over 1 -
\omega } \sqrt { ( 1 - \omega )^{ 1/\omega } \over 2 \pi \omega z^{(1
+ \omega)/ \omega } } \exp \Big [ - \omega \Big ( {z \over 1 - \omega}
\Big )^{- {{1 - \omega} \over \omega}} \Big ]; ~~~~~~~b(\delta) = -
\left ( {2 \omega \over \bar{ \xi}_2} \right )^{1/2} \left ({ 1 -
\omega \over z}  \right )^{(1 - \omega)/2 \omega} 
\end{equation}

\n
To summarize, we can say that the entire behavior of the PDF
 $P(\delta)$ is
encoded in two different scaling functions, h(x) and g(z) and one can
also study the scaling properties of $b(\delta)$ in terms of the scaling
variables $x$ and $z$ in a very similar way. These
scaling functions are relevant for small and large $\delta$ behavior
of the function $p(\delta)$ and $b(\delta)$. Typically the PDF
$p(\delta)$ shows a cutoff at
both large and small values of $\delta$ and it exhibits a
power-law in the middle. The power law behavior is observed when both
$g(z)$ and $h(x)$ overlap and is typical of highly non-linear
regime. With the decrease in $\bar \xi_2$ the range of $\delta$ 
for which $p(\delta)$ shows such a power law behavior decreases
finally to vanish for the case of very small variance i.e. in the
quasi-linear regime. Similarly the bias is very small and a slowly
varying function for moderately over dense objects but increases
rapidly for over dense objects which are in qualitative agreement 
with PS formalism.

\subsubsection{The Quasi-linear Regime}

In the quasi-linear regime, a similar formalism can be used to
study the PDF. However the generating function now
can be explicitly evaluated by using tree-level perturbative
dynamics (Bernardeau 1992; Bernardeau 1994). 
It is also possible to take smoothing 
corrections into account in which case one can have explicit
expression of $\omega$ in terms of the initial power spectral index $n$.
In general the parameters $k_a$ or $\omega$ charecterising VPF
or CPDF are different from there highly non-linear values. 

For the purpose of weak lensing calculations it is important to notice 
that although the generating function for the matter correlation
hierarchy for very small angular smoothing is the one from 
the highly non-linear
regime, the analytical results that are useful are the ones from
the quasi-linear regime as the variance of projected density field 
remains very small even for small smoothing scales. 

The PDF and bias now can be expressed in terms of $G_{\delta}(\tau)$
(Bernardeau 1992, Bernardeau 1994): 

\begin{figure}
\protect\centerline{
 \epsfysize = 3.truein
 \epsfbox[22 393 344 708]
 {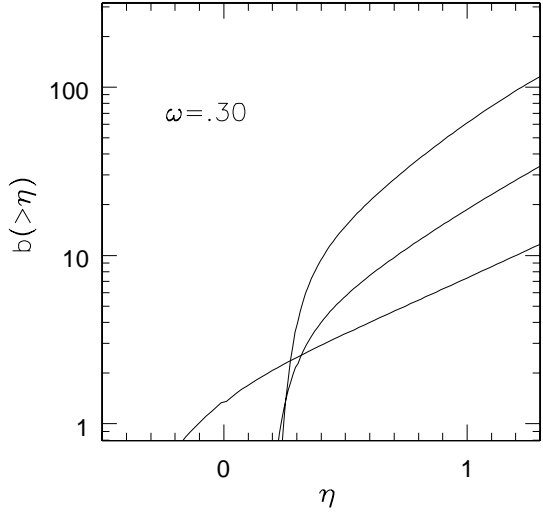} }
 \caption{The cumulative bias $b_{\eta}(>\eta)$ associated with the reduced convergence field $\eta$ is
 plotted as a function of $\eta$ for three different values of
 variance (bottom to top)  i.e. ${\sqrt \xi_{\eta}} = .25, .5,$ and $
 1$. Scaling parameter $\omega = 0.3$ is used through out out our analysis. }
\end{figure}

\begin{eqnarray}
&&p(\delta)d \delta = { 1 \over -G_{\delta}'(\tau) } \Big [ { 1 - \tau G_{\delta}''(\tau)
/G_{\delta}'(\tau) \over 2 \pi {\bar \xi}_{2} }  \Big ]^{1/2} \exp \Big ( -{ \tau^2
\over 2 {\bar \xi}_{2}} \Big ) d \tau; ~~~~~ b(\delta) = - \left (
{k_a \over \bar \xi_2} \right ) \left [ ( 1 + G_{\delta}(\tau)
)^{1/k_a} - 1 \right ] , \\
&&G_{\delta}(\tau) = G(\tau) - 1 =  \delta.
\end{eqnarray}

\n
The above expression is valid for $\delta < \delta_c$ where the $\delta_c$
is the value of $\delta$ which cancels the numerator of the pre-factor 
of the exponential function appearing in the above expression. For
$\delta > \delta_c$ the PDF develops an exponential tail which is 
related to the presence of singularity in $\phi(y)$ in a very similar
way as in the case of its highly non-linear counterpart (Bernardeau
1992, Bernardeau 1994).

\begin{equation}
p(\delta) d \delta = { 3 a_s \sqrt {{\bar \xi}_2} \over 4  {\sqrt \pi} }
\delta^{-5/2} \exp \Big [ -|y_s|{ \delta \over {\bar \xi}_{2}} + {|\phi_s|
\over {\bar \xi}_{2}} \Big ] d \delta; ~~~~~b(\delta) = -{ 1 \over
{\cal G}'(\tau_s)} {(1 + \delta)
\over { {\bar \xi}_2}} 
\end{equation}

In recent studies it was shown that the underlying smoothed
density PDF and its convergence counterpart $p_{\kappa}(\kappa_s)$ 
are exactly
same (Munshi \& Jain 1999b) under small angle approximation. In this
paper we will extend such result to show that even the bias associated 
with the convergence map can be related to the bias associated with 
over-dense objects in a very similar manner. Such results can also be
obtained from the extended PS formalism and we plan to present details
of such analytical results elsewhere.

It may be noted that similar analytical
expressions for the PDF and bias  can also be derived for the case of
approximate dynamics 
sometime used to simulate gravitational clustering in the weakly
non-linear regime or to reconstruct the projected density maps 
from convergence maps with large smoothing angles 
(e.g. Lagrangian perturbation theory which
is an extension of Zeldovich approximation (Munshi et al. 1994)).

\subsection{The Bias of the Convergence Field} 

To compute the bias associated with the peaks in the convergence field we
have to fist develop an analytic expression for the generating field
$\beta(y_1, y_2)$ for the convergence field $\kappa_s$.
For that we will use the usual definition for the two-point 
cumulant correlator $C_{pq}$ for the convergence field (for a complete
treatment of statistical properties of $\kappa_s$ see Munshi \& Coles, 1999b).

\begin{equation}
C^{\kappa}_{pq} = {\langle \kappa_s(\gamma_1)^p \kappa_s(\gamma_2)^q \rangle \over  \langle k_s^2 \rangle^{p+q-2} \langle
\kappa_s(\gamma_1) \kappa_s(\gamma_2) \rangle } = C^{\kappa}_{p1}
C^{\kappa}_{q1} 
\end{equation}

\n
We will show that like its density field counterpart the
two-point generating function for the convergence field $\kappa_s$ 
can also be
expressed (under certain simplifying assumptions) as a product 
of two one-point generating functions $\beta(y)$
which can then be directly related to the bias associated with
``hot-spots''in the convergence field. 

\begin{equation}
\beta_{\kappa}(y_1, y_2) = \sum_{p,q}^{\infty} {C^{\kappa}_{pq} \over p! q!} y_1^p y_2^q = 
\sum_{p}^{\infty} {C^{\kappa}_{p1} \over p!} y_1^p \sum_{q}^{\infty} {C^{\kappa}_{q1}
\over q!} y_2^q  = \beta_{\kappa}(y_1) \beta_{\kappa}(y_2)\equiv
\tau_{\kappa}(y_1) \tau _{\kappa}(y_2)
\end{equation}

\n
It is clear that the factorization of generating function actually
depend on the factorization property of the cumulant correlators i.e.
$C^{\eta}_{pq} = C^{\eta}_{p1} C^{\eta}_{q1}$. Note that such a factorization is 
possible when the correlation of two patches in the directions 
$gamma_1$ and $\gamma_2$ $\two$  is smaller compared to the variance
$\one$ for the smoothed patches

\begin{equation}
\beta_{\kappa}(y_1, y_2) = \sum_{p,q}^{\infty} {1 \over p! q!} { y_1^p
y_2^q\over
\langle \kappa_s^2 \rangle^{p+q-2} } {\langle \kappa_s(\gamma_1)^p \kappa_s(\gamma_2)^q \rangle  \over \langle
\kappa_s(\gamma_1) \kappa_s(\gamma_2)\rangle }.
\end{equation}

\n
We will now use the integral expression for cumulant correlators
(Munshi \& Coles 1999a) to
express the generating function which in turn uses the hierarchical 
{\em ansatz} and the far field approximation as explained above

\begin{eqnarray}
\beta_{\kappa}(y_1, y_2) &=& \sum_{p,q}^{\infty} {C^{\eta}_{pq} \over p! q! } { 1 \over 
\langle \kappa_s^2 \rangle^{p+q -2}} { 1 \over \langle
\kappa_s(\gamma_1) \kappa_s (\gamma_2) \rangle } \nonumber \\
&& \times  \int_0^{\chi_s} d\chi
{ \omega^{p+q} \over r^{2(p+q -1)} } \Big [ \int  {d^2 {\bf l} \over (2
\pi)^2 )} P( { \bf l \over r(\chi) })  W^2(l\theta_0) \exp [ i l
\theta_{12} \Big ]
\Big [ \int  {d^2 {\bf l} \over (2
\pi)^2 )}  P( { \bf l \over r(\chi) })  W^2(l\theta_0) \Big ]^{p+q-2} y_1^p y_2^q.
\end{eqnarray}

\n
It is possible to further simplify the above expression by separating the
summation over dummy variables $p$ and $q$, which will be useful to
establish the factorization property of two-point generating function
for bias $\beta(y_1, y_2)$.

\begin{eqnarray}
\beta_{\kappa}(y_1, y_2) &=& \inc d\chi  {\left ({1 \over r^2(\chi)}
\right ) \corr \over \two } {\one^2 \over{\left ({1 \over r^2(\chi)}\right )^2\var^2}
} \nonumber  \\ &&
\times \sum_{pq}^{\infty}       
 {C^{\eta}_{pq} \over p! q!} \Big ( { y_1 \over \one} {\omega(\chi) \over r^2
(\chi)} \var  \Big )^p \Big ( { y_2 \over \one} {\omega(\chi) \over r^2
(\chi) } \var \Big )^q
\end{eqnarray}

\n
We can now decompose the double sum over the two indices into two
separate sums over individual indices. Finally using the definition of
the one-point generating function for the cumulant correlators 
we can write: 

\begin{eqnarray}
\beta_{\kappa}(y_1, y_2) &=& \inc d\chi  {{\left ({1 \over r^2(\chi)}\right )} \corr \over \two } { \one^2
\over {\left ({1 \over r^2(\chi)}\right )^2\var^2}
} \nonumber  \\ &&
\times \beta_{\eta} \Big ( { y_1 \over \one} {\omega(\chi) \over r^2
(\chi)} \var  \Big )  \beta_{\eta} \Big ( { y_2 \over \one} {\omega(\chi) \over r^2
(\chi)} \var  \Big ).
\end{eqnarray}

\begin{figure}
\protect\centerline{
 \epsfysize = 1.5truein
 \epsfbox[4 4 377 194]
 {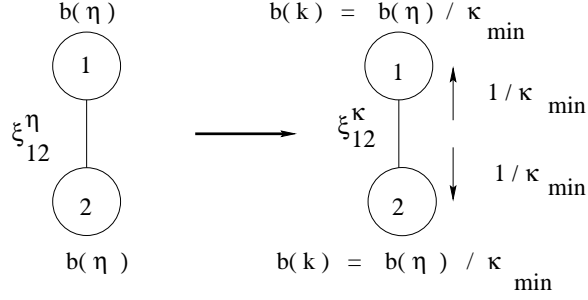} }
 \caption{The construction of the bias function $b(\kappa_s)$ for the smoothed convergence
 field from the bias $b(\eta_s)$ associated with the smoothed reduced convergence
 field $\eta_s$.}
\end{figure}

\n
The above expression is quite general and depends only on the small angle 
approximation and the large separation approximation and is valid for any
given specific model for the generating function $G(\tau)$. However it is easy
to notice that the projection effects as encoded in the line of sight
integration do not allow us to write down the two-point generating
function $\beta_{\kappa}(y_1, y_2)$ simply as a product of two one-point generating functions $\beta_{\eta}(y)$ as
was the case for the density field $1+ \delta$.  

As in the case of the derivation of the probability distribution
function for the smoothed convergence field $\kappa_s$ it will be much
easier if we define a reduced smoothed convergence field $\eta_s$. The
statistical properties of $\eta_s$ are very similar to that of the 
underlying 3D density field (under certain simplifying approximation) and are roughly
independent of the background geometry and dynamics of the universe,

\begin{equation}
\eta_s = {\kappa_s - \kappa_{min} \over -\kappa_{min}} = 1 + {\kappa_s
\over |\kappa_{min}|}.
\end{equation}

\n
Where the minimum vale of $\kappa_s$ i.e. $k_{min}$ is defined as:

\begin{equation}
k_{min} = -\int_0^{\chi_s} d \chi \omega (\chi).
\end{equation}

\n
It is easy to notice that the minimum value of the convergence field
will occur in those line of sight which are completely
devoid of any matter i.e. $\delta = -1 $ all along the line of
sight. We will also find out later that the
cosmological dependence 
of the statistics of $\kappa_s$ field is encoded in $k_{min}$ and 
this choice of the new variable $\eta_s$ makes its related 
statistics almost
independent of the background cosmology. Repeating the above analysis
again for the $\eta_s$ field,  we can 
express the cumulant correlator generating function for the reduced 
convergence field $\eta_s$ as:

\begin{eqnarray}
\beta_{\eta}(y_1, y_2) &=& \inc \omega^2(\chi) d\chi  {{\left ( \omega^2(\chi) \over
r^2(\chi) \right )} \corr \over \two } { \one^2 \over
{\left ( \omega(\chi) \over
r^2(\chi) \right )^2} \var^2 \Big [ \kmin \Big ]^2
} \nonumber  \\ &&
\times \beta_{\eta} \Big ( { y_1 \over \one} {\omega(\chi) \over r^2
(\chi)} \var \kmin  \Big ) \nonumber \\ &&
\times  \beta_{\eta} \Big ( { y_2 \over \one} {\omega(\chi) \over r^2
(\chi)} \var  \kmin \Big ).
\end{eqnarray}

While the above expression is indeed very accurate and relates the
generating function of the density field with that of the convergence
field, it is difficult to handel for any practical purpose.
Also it is important to notice that the scaling functions such 
as $h(x)$ for the 
density probability distribution function and  $b(x)$ for 
the bias associated with over-dense objects are typically estimated
from numerical simulations specially in the highly non-linear regime. 
Such estimations are plagued with several uncertainties such as
finite size of the simulation box. It was noted in earlier studies 
that such uncertainties lead to only a rather approximate estimation
of $h(x)$. The estimation of the scaling function associated with the
bias i.e. $b(x)$ is even more complicated due to the fact that the
two-point quantities such as the cumulant correlators and the
bias are more affected by finite size of the catalogs. 
So it is not  fruitful to 
actually integrate the exact integral expression we have derived
above and we will replace all line of sight integrals with its
approximate values. recent study by Munshi \& Jain (1999) have used an exactly similar
approximation to simplify the one-point probability distribution
function for $\kappa_s$ and found good agreement with ray tracing
simulations. We will show that our approximation reproduces the 
numerical results quite accurately for a wide range of smoothing angle,

\begin{eqnarray}
&& |\kappa_{min}| \approx {1\over 2} \chi_s \omega(\chi_c), \\ 
&&\one  \approx {1\over 2} \chi_s  \omega^2(\chi_c) \Big [ {d^2 k \over
(2\pi)^2} {\rm P(k)} W^2(k r(\chi_c) \theta_0) \Big ],\\ 
&&\two  \approx {1\over 2} \chi_s  \omega^2(\chi_c) \Big [ {d^2 k \over
(2\pi)^2} {\rm P(k)} W^2(k r(\chi_c) \theta_0) \exp [i k r(\chi_c) \theta_{12}] \Big ].
\end{eqnarray}

\n
Use of these approximations gives us the leading order contributions
to these integrals and we can check that to this order we recover the 
factorization property of the generating function i.e. $\beta_{\eta}(y_1,
y_2) = \beta_{\eta}(y_1) \beta_{\eta}(y_2)$,

\begin{equation}
\beta_{\eta}(y_1, y_2) = \beta_{\eta}(y_1) \beta_{\eta}(y_2) =
\beta_{1+\delta}(y_1) \beta_{1+\delta}(y_2) \equiv \tau(y_1)\tau(y_2).
\end{equation}

So it is clear that at this 
level of approximation, due to the factorization property of the cumulant 
correlators, the bias function $b_{\eta}(x)$ associated with the peaks
in the convergence field $\kappa_s$,  beyond certain threshold, obeys a 
similar factorization property too, which is exactly same as its
density field counterpart. Earlier studies have established such a 
correspondence between convergence field and density field 
in the case of one-point probability distribution function $p(\delta)$
(Munshi \& Jain 1999b),

\begin{equation}
b_{\eta}(x_1) h_{\eta}(x_1) b_{\eta} (x_2) h_{\eta} (x_2) = 
 b_{1 + \delta}(x_1) h_{1 + \delta}(x_1) b_{1+\delta} (x_2) h_{1+\delta} (x_2).
\end{equation}

\n
Where we have used the following relation between $\beta_{\eta}(y)$
and $b_{\eta}(x)$,

\begin{equation}
b_{\eta}(x) h_{\eta}(x) = -{ 1 \over 2 \pi i}
\int_{-i\infty}^{i\infty} dy \tau (y) \exp (xy).
\end{equation}

\n
For all practical purpose we found that the differential bias 
as defined above is more difficult to measure from numerical
simulations as compared to its 
integral counterpart where we concentrate on the bias associated with
peaks above certain threshold,

\begin{equation}
b_{\eta}(>x) h_{\eta}(>x) = -{ 1 \over 2 \pi i}
\int_{-i\infty}^{i\infty} dy {\tau (y)\over y} \exp (xy).
\end{equation}

\n
It is important to notice that although the bias $b(x)$
associated with the convergence field and the underlying density
field is exactly same, the variance associated with the density field is
very high but the projection effects in the convergence field 
brings down the variance in the convergence field to less than unity
which indicates that we have to use
the integral definition of bias to recover it from its generating
function (see eq.(\ref{ber1}) and eq.(\ref{ber2})). 

Now writing down the full two point probability distribution function 
for two correlated spots in terms of the convergence field $\kappa$ and
its reduced version $\eta$:

\begin{eqnarray}
&&p_{\kappa}(\kappa_1, \kappa_2)d\kappa_1 d\kappa_2 = p_{\kappa}(\kappa_1) p_{\kappa}(\kappa_2)( 1
+ b_{\kappa}(\kappa_1) \xi^{\kappa}_{12} b_{\kappa}(\kappa_2)) d\kappa_1 d\kappa_2, \\ 
&&p_{\eta}(\eta_1, \eta_2)d\eta_1 d\eta_2 = p_{\eta}(\eta_1) p_{\eta}(\eta_2)( 1
+ b_{\eta}(\eta_1) \xi^{\eta}_{12} b_{\eta}(\eta_2)) d\eta_1 d\eta_2  
\end{eqnarray} 

\n
In our earlier analysis (Munshi et al. 1999b) we found that $p^{\kappa}(\kappa)
= {p^{\eta}(\eta) \over {k_{min}}}$ we also noticed that $\xi^{\kappa}_{12} = 
{\xi^{\eta}_{12}\over {\kappa_{min}}^2}$.
Using these relations we can now write:

\begin{equation}
b_{\kappa}(\kappa) = {b_{\eta}(\eta) \over {k_{min}}}.
\end{equation}

\n
This is one of the main result of our analysis and in the next section we
will show that it is indeed a very good approximation to numerical 
simulations.

\begin{figure}
\protect\centerline{
 \epsfysize = 4.truein
 \epsfbox[22 147 583 708]
 {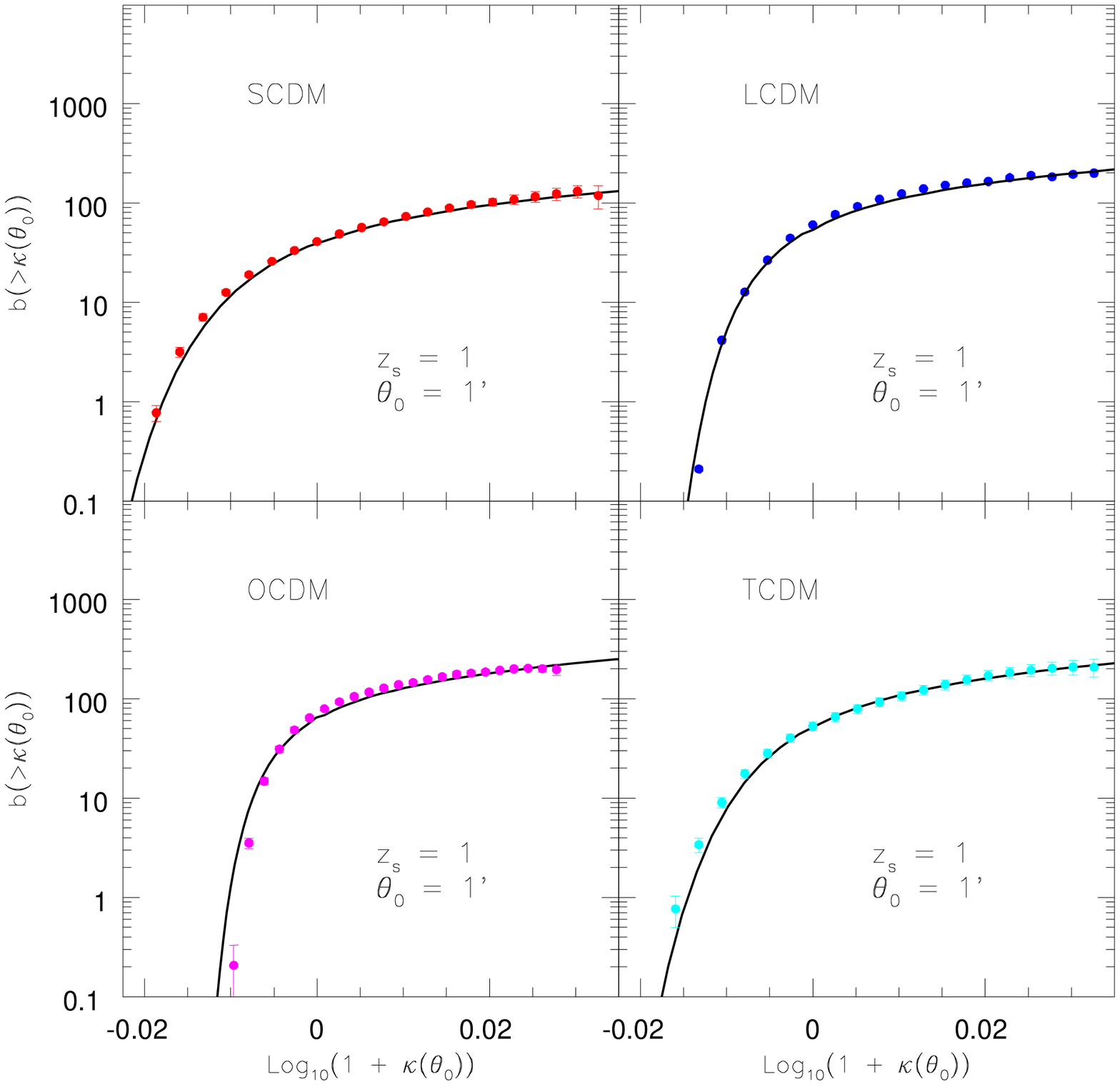} }
 \caption{The cumulative bias function $b(>\kappa(\theta_0))$ is
 plotted against the smoothed convergence field
 $\kappa(\theta_0)= \kappa_s$.
The different panels correspond to the different 
cosmological models and the smoothing angle $\theta_0$ is 
fixed at $1'$. The solid line at each panel correspond to analytical
 predictions from the hierarchical {\em ansatz}. The source redshift
 $z_s$ is fixed at unity.}

\end{figure}

\begin{figure}
\protect\centerline{
 \epsfysize = 4.truein
 \epsfbox[22 147 583 708]
 {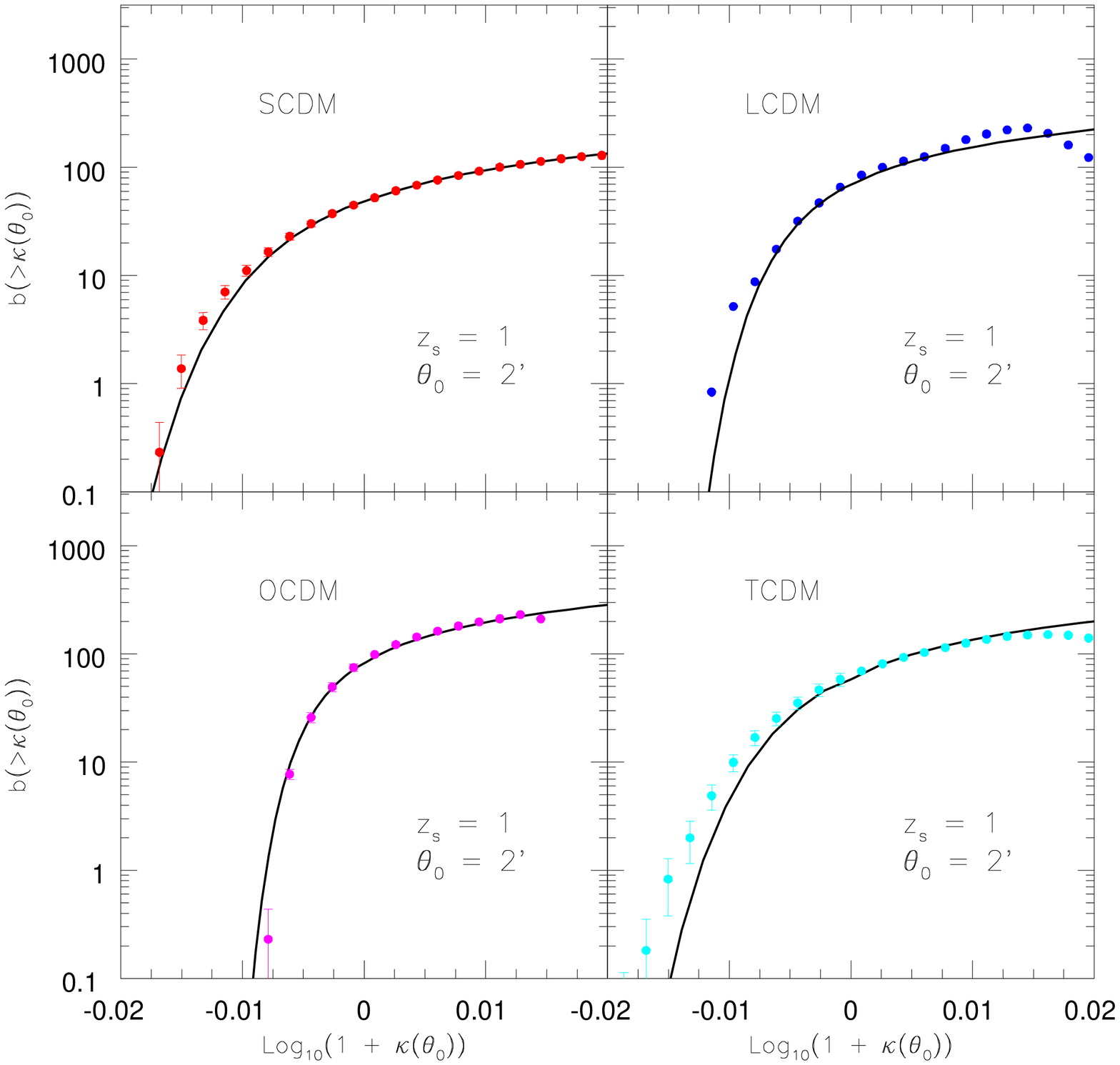} }
 \caption{Same as Figure - 3 but for smoothing angle  $\theta = 2'$.}
\end{figure}

\begin{figure}
\protect\centerline{
 \epsfysize = 4.truein
 \epsfbox[22 147 583 708]
 {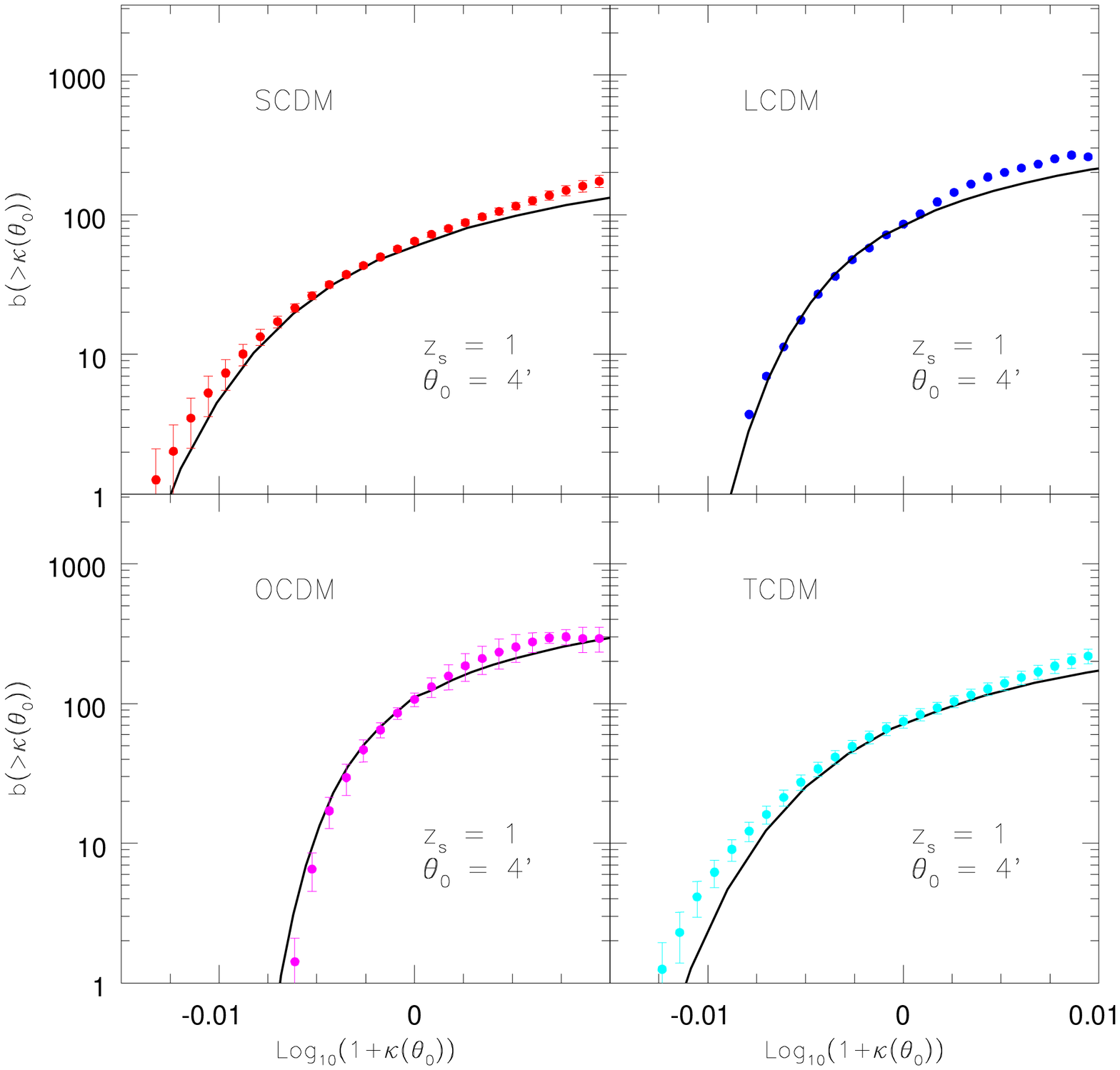} }
 \caption{Same as Figure - 3 but for smoothing angle  $\theta = 4'$.}
\end{figure}

\begin{figure}
\protect\centerline{
 \epsfysize = 4.truein
 \epsfbox[22 147 583 708]
 {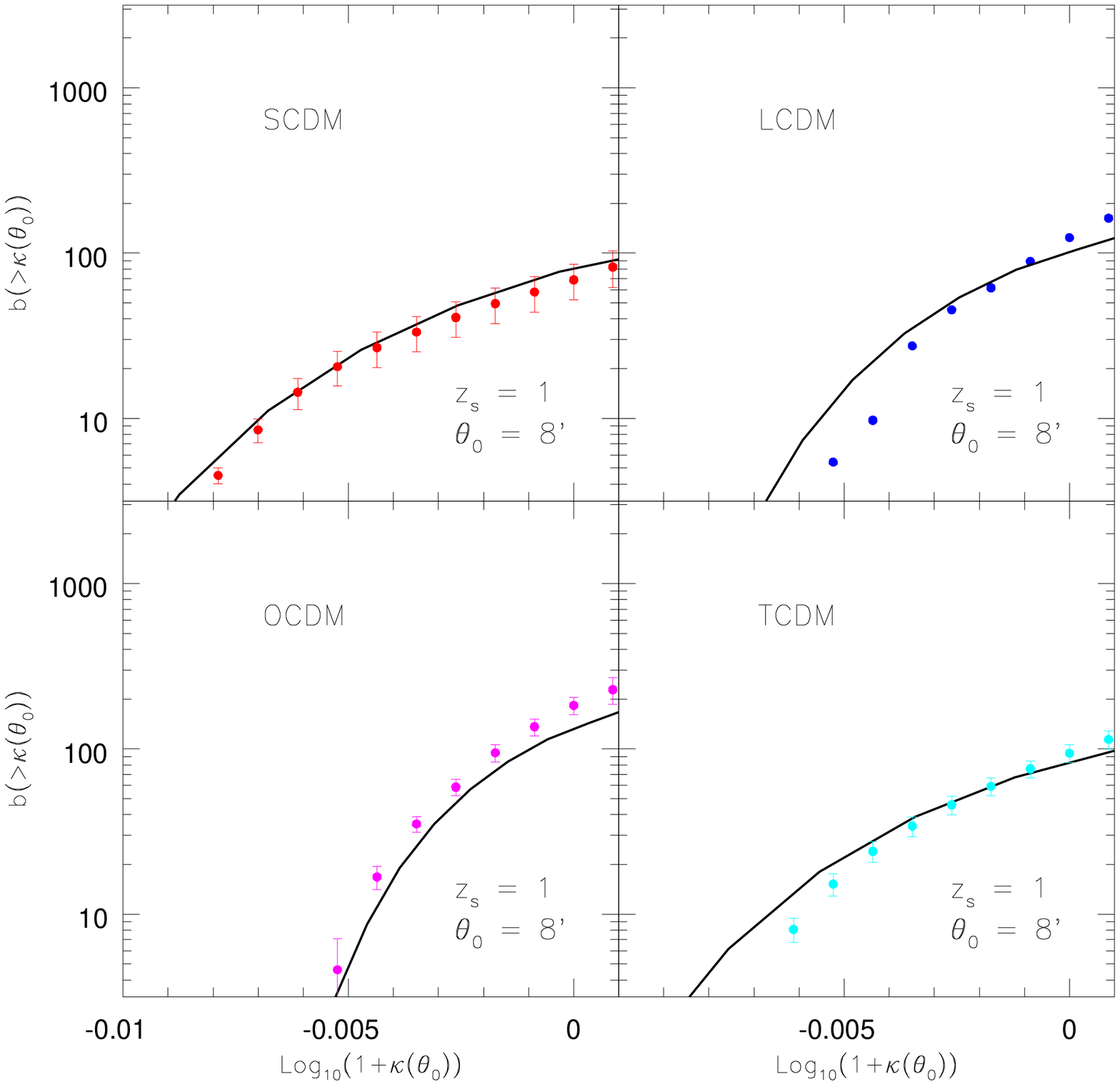} }
 \caption{Same as Figure - 3 but for smoothing angle  $\theta = 8'$.}
\end{figure}


\begin{figure}
\protect\centerline{
 \epsfysize = 4.5truein
 \epsfbox[19 144 586 714]
 {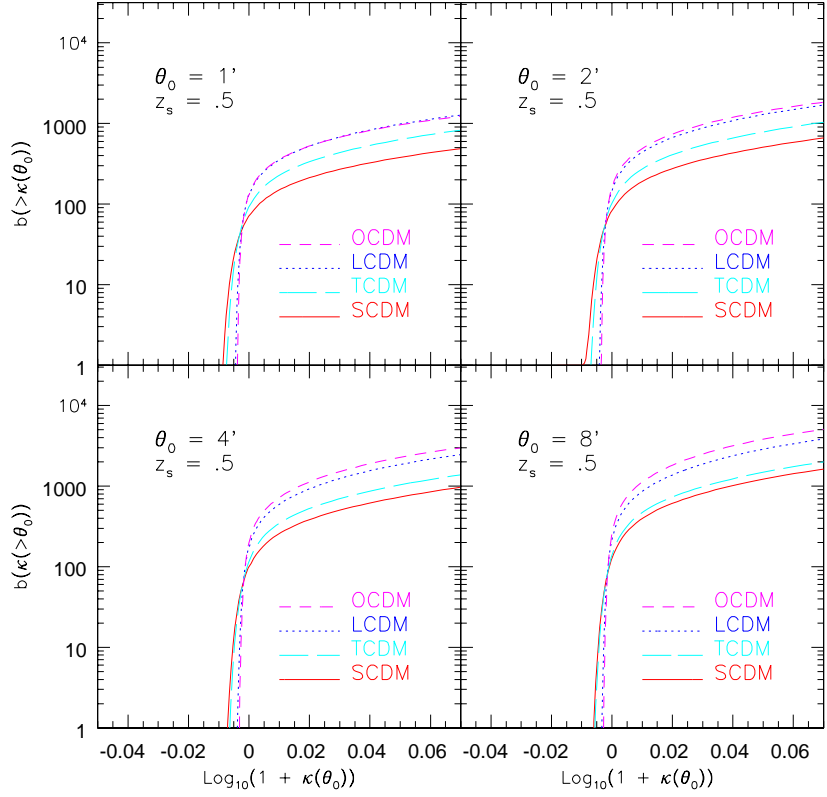} }
 \caption{The analytical predictions for the bias associate with
the convergence field $\kappa(\theta_0)$ is plotted as a function of 
$\kappa(\theta_0)$. The smoothing angle varies from 1' to 
8' as indicated. The source redshift is fixed at $z_s = .5$. Different
curves correspond to SCDM, LCDM, OCDM and TCDM models.
}
\end{figure}

\begin{figure}
\protect\centerline{
 \epsfysize = 4.5truein
 \epsfbox[19 144 586 714]
 {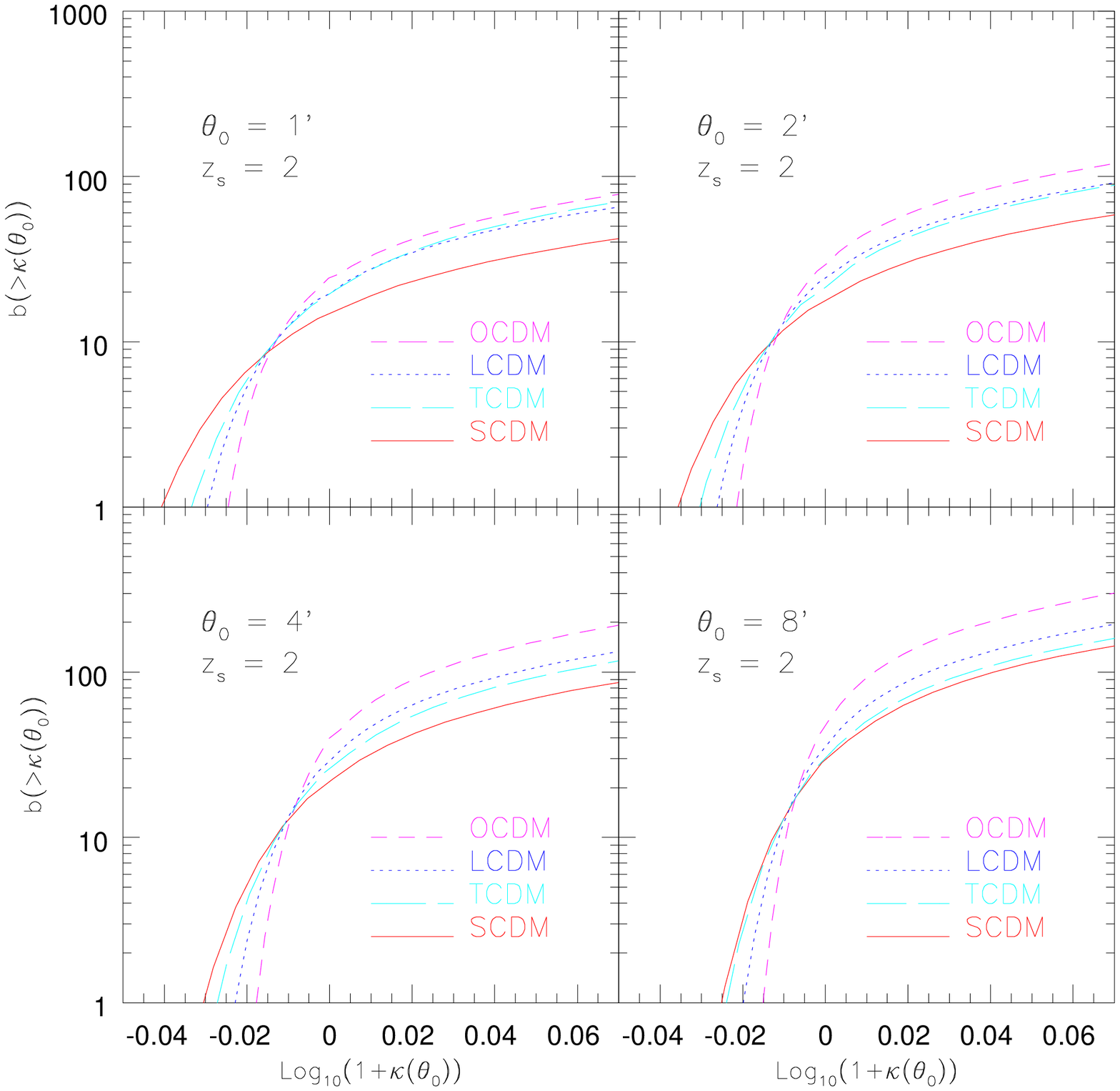} }
 \caption{Same as Figure - 7 but for the redshift $z_s = 2$. Note that the
 bias associated with high $\kappa(\theta_0)$ spots in the convergence
 field decreases as we increase the source redshift $z_s$.}
\end{figure}

\section{Comparison against Numerical Ray Tracing Simulations}

To compare the analytical results with numerical simulations we
have smoothed the convergence field or $\kappa$- map generated from numerical
simulations using a top-hat filter of suitable smoothing angle $\theta_0$.
The minimum smoothing radius we have used is $1'$ which is 
much larger compared to the numerical resolution length scale. In our 
earlier studies we found that numerical artifacts are negligible
even for angles as small as $.25'$. The maximum 
smoothing radius we have studied is $8'$, which is much 
smaller compared to the box size of all models and we expect 
that the finite volume corrections are
not significant in our studies. The box size is L = 166.28' for the EDS models,
L =235.68' for $\Omega=0.3$ open model and L = 209.44' for 
$\Omega=0.3$  model with cosmological constant $\Lambda =
0.7$. The numerical outputs from more than ten realizations of 
these cosmological models were used to find the average and the 
scatter and were tested against the analytical predictions. 
For LCDM models we
have used only one particular relaisation.

For the determination of the bias from the convergence maps we found
that the computation
of cumulative bias $b(>\kappa_s)$ is much  more stable compared to the
computation of the differential bias $b(\kappa_s)$. We have 
therefore used the analytical predictions for $b(>\kappa_s)$ to
compare
against numerical simulations. The bias we have studied will
represent the correlation of those spots in the convergence maps which 
are above certain threshold and is defined as:
 

\begin{equation}
b_{\kappa}(> \kappa_s) = {1 \over \sqrt {\two}} {\left ({\int_{\kappa_s}^{\infty}d \kappa_1
\int_{\kappa_s}^{\infty}d \kappa_2 { p_{\kappa}(\kappa_1, \kappa_2)} \over
\left [ \int_{\kappa_s}^{\infty} p_{\kappa}(\kappa_1) d \kappa_1 \right ]^2} - 1  \right )}^{1/2}.
\end{equation}

\n
The above expression relies on evaluating bias by finding out directly
correlation of points in the convergence map which are above certain
threshold and does not depend on any factorization property of bias.
%
%
%
We have  tested the functional form for the bias as predicted from
the analytical model against numerical
simulations. The analytical prediction tells us that the bias 
associated with
spots where the smoothed convergence field is negative is 
very small and is almost zero. On the other hand the negative
 convergence spots are not
biased at all, this property also holds for the bias
associated with under-dense regions in the density field which
are responsible for producing the negative spots in the convergence field. For positive 
peaks in the convergence map the bias is positive and it increases sharply
as we move towards higher thresholds for $\kappa_s$. Contributions to
positive spots in the convergence maps comes from density peaks in 
underlying 3D mass distributions and again its remarkable that the
bias associated with both of them are actually described by the same
functional form. While we change variables from the reduced convergence
$\eta_s$ to the actual convergence field $\kappa_s$ we introduce a 
multiplicative factor of $1/k_{min}$. As we explained before that 
this term introduces the cosmological dependence for the bias in the 
convergence field. This multiplicative factor is largest for the OCDM
model and is smallest for the SCDM model and hence the bias 
associated with peaks in the convergence maps also follow the same 
pattern.

The analytical results and results of numerical simulations are 
plotted in  
figures - 3-6. Comparing the analytical results with the results
obtained from the numerical
simulations we find that these predictions match very accurately. The general trend of
increase of bias with $\kappa_s$ is well reproduced by the analytical
models. Also we find that the analytical model is able to describe the
relative ordering of the bias function for different cosmological
models quite accurately. We have also computed the scatter in 
numerical results and found them to increase with $\kappa_s$. With
increasing the threshold we decrease the number of spots which cross
that
threshold thereby increasing the sample variance. The sample variance also
increases with the smoothing length scale $\theta_0$ due to decrease
in the number
of smoothed patches which carry completely independent information.
To improve upon the situation we have increased the sampling by
shifting the grid used to bin the convergence map in two perpendicular
direction. Although there has been several detailed studies 
to quantify  
the effect of finite volume correction on the statistics of galaxy
clustering, similar studies are still lacking for weak lensing
statistics. However it was found in numerical studies 
clustering that the computed PDF shows a sharp cutoff at some maximum 
value of cell count $n_{max}$ which is the densest cell in the
catalog. Increasing the size of the catalog will increase $n_{max}$ 
up to which one can reliably compute the PDF. The long exponential 
tail of PDF also shows large fluctuations due to the presence or absence
of any rare dense objects in the N-body catalog. In this paper we have
shown how intimately  statistics of the weak lensing convergence field
$\kappa_s$ and its density field counterpart are related. This will 
indicate that the numerical computations of the one-point PDF and 
the bias of 
$\kappa_s$ and $\delta$ are both affected in a very similar manner due
to the finite size of the catalog. It is therefore quite remarkable
that despite the fact that we have not introduced any involved correction procedure in our measurement of bias or PDF the
numerical results match the analytical predictions so accurately.
We have not studied the factorization property of the bias as
predicted by hierarchical {\em ansatz}, as there has not been any
parallel study in this direction so far dark matter clustering in 3D.
However we plan to extend our results to such analysis elsewhere.

We have studied both analytically and numerically
 how the bias changes if we 
change the source redshift $z_s$. In our earlier studies of PDF we 
have shown that for low source redshift when one probes the highly
non-linear regime directly, the PDF is charecterised by a long
exponential tail originating from collapsed objects. Where as if we 
increase the source redshift we are adding more slabs of matter in
the intervening path which are virtually uncorrelated and hence it 
makes the distribution more Gaussian. However it also increases the
variance of the convergence field. We find a similar result for the 
bias in $\kappa_s$ field too, for low source redshift the highly 
dense spots which are contributing towards the nearby  high $\kappa_s$spots
may also be physically be very near by and hence they themselves may
be correlated more strongly than the underlying mass distribution. 
On the
other hand when we increase the source redshift we increase the number
of hight density spots which will produce the high convergence
$\kappa_s$ spots in the map, however it is to be kept in mind that
clearly not all these high density spots are in physical proximity in 
3D space and might exist in different slabs of the underlying mass
distribution. This explains why although the variance of $\kappa_s$
and correlation in $\kappa_s$ do increase with the source redshift 
but the bias associated with high $\kappa_s$ spots actually decrease
with the source redshift.

In our analytical studies we have assumed that the small angle
approximation is valid and we also assumed that the separation angle
$\theta_{12}$ is much larger than the smoothing angle $\theta_0$. 
This will imply that
variance within the smoothed patch is much larger compared to the correlation 
between the two smoothed patches. However the separation angle should
again be small so that we can still replace all spherical harmonics by
Fourier modes. In numerical simulations we have placed the smoothed 
patches separated by an angle at least by thrice the smoothing angle. 
In almost all cases the separation angle is 3.5 times larger than the 
smoothing angle. We have also found (as predicted by analytical 
results) that once the large separation limit is reached the bias 
function no-longer depends on the separation angle $\theta_{12}$.

\section{Discussion}

The hierarchical {\em ansatz} provides a framework in which 
gravitational clustering is generally studied in the highly non-linear
regime. However
most of such earlier studies are directed towards understanding of 
galaxy clustering and were tested against numerical N-body
simulations. However studies based on clustering properties of
galaxies are more difficult to interpret as they always depend on
a particular biasing scheme employed to relate galaxy population to
the underlying mass fluctuation. Forthcoming weak lensing surveys will 
provide an unique opportunity to directly study matter distribution 
in the universe and will help us in understanding their clustering
properties. The main motivation of this paper 
is to extend use of hierarchical {\em ansatz} to the study of bias 
as measured from weak lensing surveys.

There have been several studies in which perturbative techniques
were used to relate the convergence statistics measured from 
weak lensing 
surveys to the clustering statistics of the underlying mass 
distribution.
However analytical predictions based on perturbative techniques 
normally requires a large smoothing angle. Most of early observational
studies of weak lensing surveys will focus on smaller angular scales
and hence will require a completely non-linear treatment. In this 
paper we have shown that our current level of understanding of 
gravitationally clustering is enough to make concrete predictions
for two-point statistics of the convergence field.

We have used the non-local scaling relations to evolve the matter 
two-point correlation function and used the hierarchical {\em ansatz} 
to relate
the higher order correlation function with two-point correlation
function. Combined with the generating function approach these two {\em ansatze}
provide a powerful tool for the analysis of the weak lensing maps and its
multi-point statistics. In accompanying papers we have computed the 
lower order cumulants and the cumulant correlators for various smoothing
angle and for different source redshifts (Munshi \& Jain 1999a). 
We have also studied the complete PDF for the smoothed convergence field 
(Munshi \& Jain 1999b). In this paper we have extended these studies 
by testing analytical predictions for the bias against numerical simulations.
The cumulant correlators are nothing but lower order moments of the
bias function which we have 
studied in this paper. We found a very good agreement between analytical
results and ray tracing simulations.

The bias which we have studied here is induced by gravitational
clustering. It is known from earlier studies that such a bias occurs 
naturally both in the quasi-linear (Bernardeau 1994) and the highly non-linear
regime for over-dense objects (Bernardeau \& Schaeffer 1992, 1999,
Munshi et al. 1999a,b). 
It is factorizable and will only depend on the scaling 
parameters $x$ associated with collapsed objects. We have extended such analytical
studies for the weak-lensing convergence field and have shown
that similar results do indeed hold even for such cases.  Our numerical study
confirms such analytical claims and also shows that the functional
form for the bias
associated with the convergence field is exactly same as that for 
over-dense spots in underlying mass distribution, so measurement of 
the bias from weak lensing observations will provide a direct probe 
into the 
bias associated with over-dense dark objects in the underlying mass
distribution as induced by gravitational clustering. This will help
us to separate the contribution to the bias due to gravitational
clustering from the bias associated with other non-gravitational
sources involved in galaxy formation process.

Although our formalism (which is  based on formalism developed by
Bernardeau \& Schaeffer (1992)) relates the 
bias of overdense cells in density field with ``hot-spots'' of the 
convergence map. Similar results can be obtained by using the 
extended Press-Schechter formalism to relate the peaks in the 
convergence field directly with the
bias associated with collapsed halos. A detailed analysis will be 
presented elsewhere.

The hierarchical {\em ansatz} not only provides the two-point
correlation function (i.e. the bias which we have studied here)
 for the over-dense objects but it also provides a recipe
for computing the higher order correlation functions and the 
associated
cumulants and cumulant correlators. Our formalism which we have
developed here can directly be extended to compute such quantities 
in the highly non-linear regime i.e. at small smoothing angle.
 However a detailed error analysis is necessary to study the
 possibility of
estimating such quantities from weak lensing observations.

Our analytical results use the top-hat filter, however it 
was pointed out before that the compensated filter is more suitable
for 
observational studies. We plan to extend our analytical and numerical 
studies for other filters and results will be presented elsewhere.
However it should be kept in mind that although nearby cells do 
have correlated convergence fields after smoothing with the top-hat 
window, such correlations are much less significant for the 
compensated
filter (a property which makes it more suitable for the observational 
purposes for PDF studies).

It is clear that the Born approximation is 
a necessary ingredient in the analytical computation which we have
undertaken in this paper. The Born approximation neglects all 
the higher order
terms in the photon propagation equation. It is then obvious that any 
statistical quantity such as the higher order cumulants and the cumulant
correlators will have correction terms which are neglected in our approach.
In the quasi-linear regime i.e. for large smoothing length scales 
these correction terms can in fact be evaluated using perturbative
calculations. However such an approach is not possible to adopt in the
highly non-linear regime,
because in small angular scales where most contribution comes from
length scales with variance well above unity the 
whole perturbative analysis breaks down and the only way we can check
such an approximation is to compare analytical predictions against
numerical simulations. Earlier studies of lower order 
cumulant correlators confirmed that such correction terms are indeed
negligible even in small smoothing angle (Munshi \& Jain 1999b).
A very good match that we have found for the bias function in this 
study shows that such correction terms
are indeed negligible at arbitrary order.

\section*{Acknowledgment}
I was supported by a fellowship from Humboldt foundation at MPA
when this work was completed. It is pleasure for me to acknowledge
many helpful discussions with Bhuvnesh Jain, Francis Bernardeau,
Patrick Valageas, Peter Coles and Adrian
L. Melott. The complex integration routine I have used to generate 
$b(>\eta)$ was made available to me by Francis Bernardeau. I am
grateful to him for his help. The ray tracing simulations were carried
out by Bhuvnesh Jain. I am greatly indebted to him for allowing me to 
use the data which made the present study possible.

\end{document}